\pdfoutput=1
\documentclass[runningheads]{llncs}
\usepackage[T1]{fontenc}
\usepackage{graphicx}
\usepackage{subfigure}
\usepackage{cite}
\usepackage{graphicx}
\usepackage{textcomp}
\usepackage{subfigure}
\usepackage{blindtext}
\usepackage[namelimits]{amsmath} %数学公式
\usepackage{amssymb}             %数学公式
\usepackage{amsfonts}            %数学字体
\usepackage{mathrsfs}            %数学花体
\usepackage{booktabs}
\usepackage{multirow}
\usepackage[misc]{ifsym}
\begin{document}
\title{ULTRA: Uncertainty-aware Label Distribution Learning for Breast Tumor Cellularity Assessment}
\titlerunning{Label Distribution Learning for Tumor Cellularity Assessment}

%\author{Xiangyu Li\inst{1}\orcidID{0000-0001-7681-7787} \and
%	Xinjie Liang\inst{1}\and
%	Gongning Luo\inst{1}\orcidID{0000-0003-3662-0335} \and
%	Wei Wang\inst{1}\orcidID{0000-0002-1874-9947} \and
%	Kuanquan Wang\inst{1}\orcidID{0000-0003-1347-3491} \and
%	Shuo Li\inst{2}\orcidID{0000-0002-5184-3230}} 
\author{Xiangyu Li\inst{1} \and
	Xinjie Liang\inst{1}\and
	Gongning Luo \inst{1} \textsuperscript{(\Letter)} \and
	Wei Wang\inst{1} \and
	Kuanquan Wang\inst{1}\textsuperscript{(\Letter)} \and
	Shuo Li\inst{2}} 
% index {Li,Xiangyu} {Liang,Xinjie} {Luo,Gongning} {Wang,Wei} {Wang,Kuanquan} {Li,Shuo}
\institute{Harbin Institute of Technology, Harbin, China \\
		\email{\{luogongning,wangkq\}@hit.edu.cn} \and
	Western University, London, ON, Canada\\
	%Digital Imaging Group of London, London, ON, Canada\\
	}

\authorrunning{X. Li. et al.}
\maketitle
\begin{abstract}
Neoadjuvant therapy (NAT) for breast cancer is a common treatment option in clinical practice. Tumor cellularity (TC), which represents the percentage of invasive tumors in the tumor bed, has been widely used to quantify the response of breast cancer to NAT. Therefore, automatic TC estimation is significant in clinical practice. However, existing state-of-the-art methods usually take it as a TC score regression problem, which ignores the ambiguity of TC labels caused by subjective assessment or multiple raters. 
In this paper, to efficiently leverage the label ambiguities, we proposed an \textbf{U}ncertainty-aware \textbf{L}abel dis\textbf{TR}ibution le\textbf{A}rning (\textbf{ULTRA}) framework for automatic TC estimation. The proposed ULTRA first converted the single-value TC labels to discrete label distributions, which effectively models the ambiguity among all possible TC labels. Furthermore, the network learned TC label distributions by minimizing the Kullback-Leibler (KL) divergence between the predicted and ground-truth TC label distributions, which better supervised the model to leverage the ambiguity of TC labels. Moreover, the ULTRA mimicked the multi-rater fusion process in clinical practice with a multi-branch feature fusion module to further explore the uncertainties of TC labels.
We evaluated the ULTRA on the public BreastPathQ dataset. The experimental results demonstrate that the ULTRA outperformed the regression-based methods for a large margin and achieved state-of-the-art results. The code will be available from \url{https://github.com/PerceptionComputingLab/ULTRA}.
\end{abstract}
\section{Introduction}
Breast cancer is one of the most common cancers for women worldwide\cite{key2001epidemiology}. For breast cancer treatment, neoadjuvant therapy (NAT)\cite{thompson2012neoadjuvant}, which aims to reduce the tumor size and avoid mastectomy for patients, is a common treatment option in clinical practice \cite{loibl2015neoadjuvant, rubovszky2017recent}. 
Tumor cellularity (TC), which represents the percentage of invasive tumors in the tumor bed, has been widely used to quantify the response of breast cancer to NAT\cite{rajan2004change, kumar2014study, park2016pathologic}. Besides, TC is also a significant indicator of the residual cancer burden (RCB), which predicts cancer recurrence and patients' survival\cite{symmans2007measurement}. In the current clinical practice, TC is estimated by the pathologists on hematoxylin and eosin (H\&E)-stained slides, which is rather time-consuming and suffers from inter-rater and intra-rater variability\cite{smits2014estimation}. Therefore, it is highly desirable to develop automatic and reliable methods for TC estimation. However, it is still challenging for automatic TC estimation methods because of the textural variations of different tissue types and the tissue color variations induced by differences in the slide generation process\cite{madabhushi2016image,tizhoosh2018artificial}.

Still, there are lots of methods proposed trying to achieve accurate TC estimation in a fully automatic manner. Peikari et al.\cite{peikari2017automatic} utilized the traditional machine learning method to estimate tumor cellularity. They first conducted nuclei segmentation and malignant epithelial figures classification, then performed cellularity estimation based on the malignant images.
\begin{figure}
	\centering
	\includegraphics[height=5.919cm,width=12cm]{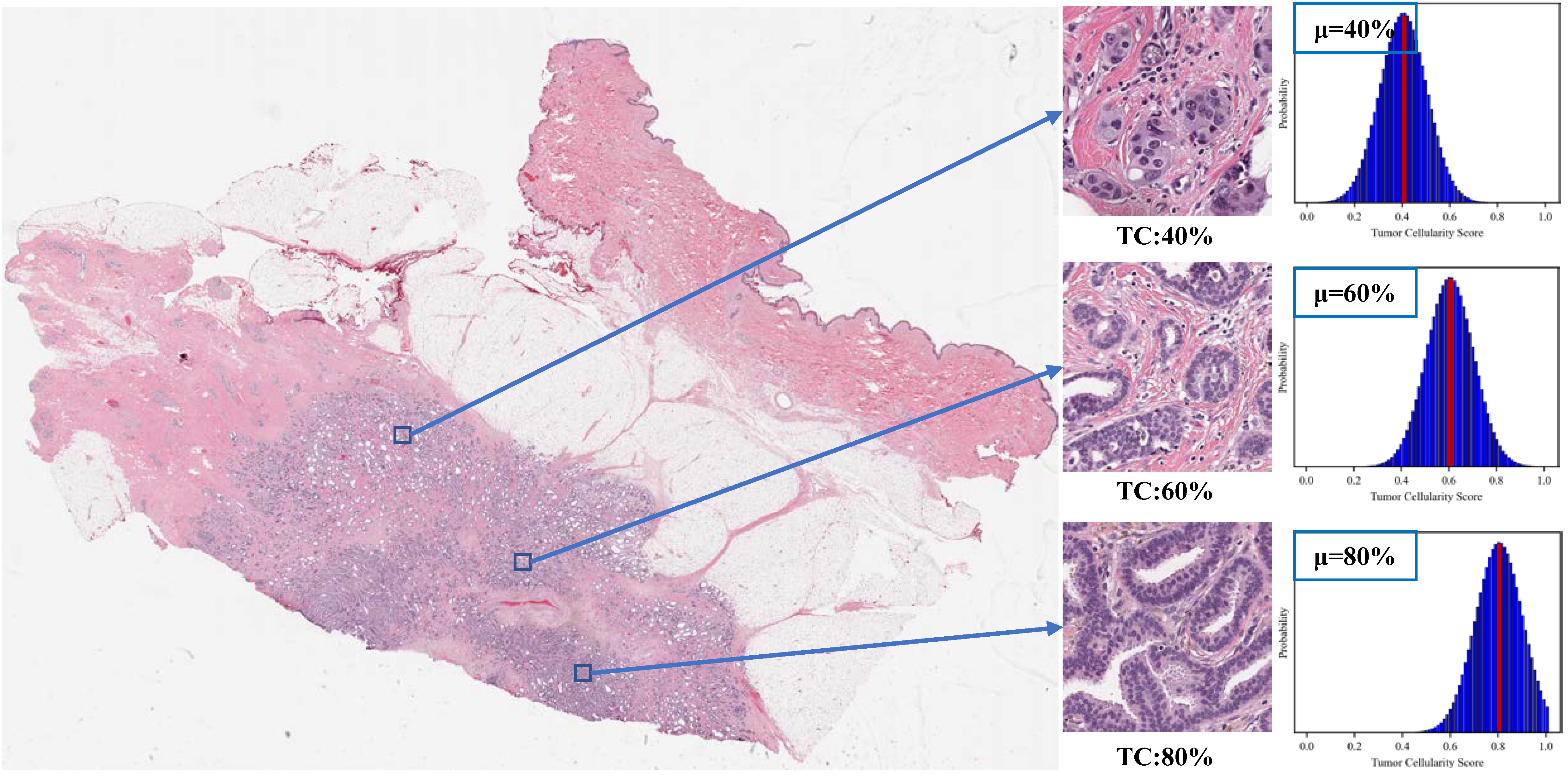}
	\caption[]{The illustration of the tumor cellularity (TC) assessment task and the core idea of the proposed ULTRA which transfers the traditional TC regression problem to a label distribution learning \cite{geng2016label} problem.} 
	\label{fig:first_figure}
\end{figure} 
Likewise, Akbar et al.\cite{akbar2019automated} proposed to conduct tumor cellularity assessment by measuring the proportion of malignant cells in the region of interest (ROI). Although the above methods have made some progress on TC estimation, they need nuclei segmentation labels which are expensive to obtain, and the performance is relatively poor. To address the above problems, Rakhlin et al.\cite{rakhlin2019breast} performed TC estimation with a cellularity score regression method, which skipped the intermediate segmentation and achieved superior results. Similarly, Akbar et al.\cite{akbar2018determining} proposed to regress tumor cellularity with ResNet\cite{he2016deep} directly. They trained a series of ResNet architectures to conduct regression tasks, and achieved better results than traditional machine learning methods.
%Srinidhi et al.\cite{srinidhi2022self} proposed a self-supervised method which take advantage of the multi-resolution contextual cues in histology WSI, and a teacher-student semi-supervised consistency paradigm that learns to effectively transfer the pretrained representations to downstream tasks based on prediction consistency with the task-specific unlabeled data.
Although those techniques achieved superior performance compared to the segmentation-based methods, they took TC estimation as a simple regression problem and ignored the intrinsic ambiguity of the TC labels caused by subjective assessment or multiple raters, which further restricted the performance improvements. 

Inspired by \cite{geng2016label,gao2017deep,tang2020uncertainty}, we proposed an \textbf{U}ncertainty-aware \textbf{L}abel dis\textbf{TR}ibution le\textbf{A}rning (\textbf{ULTRA}) framework that fully leverages the label ambiguity (uncertainty) for TC estimation. The core idea is illustrated in Fig. \ref{fig:first_figure}. The ULTRA first converted the single-value TC labels to discrete label distributions, which effectively models the ambiguity among all possible TC labels. Furthermore, the network learned the TC label distributions by minimizing the Kullback-Leibler (KL) divergence between the predicted and ground-truth TC label distributions, which better supervised the model to leverage the ambiguity of TC labels caused by subjective assessment. Moreover, the ULTRA mimicked the multi-rater fusion process in clinical routines with a multi-branch feature fusion module to further explore the uncertainties of TC labels from multiple raters. The main contributions of our paper are three-fold:
\begin{itemize}
	\item We are the first to model label ambiguity (uncertainty) of TC labels by transferring the TC score regression to a label distribution learning problem.
	\item We proposed a multi-branch feature fusion module by mimicking the multi-rater fusion process in clinical routines, which effectively leveraged the label uncertainty and significantly improved the TC estimation performance.
	\item Our ULTRA outperforms both segmentation-based and regression-based methods on the TC estimation task and achieved state-of-the-art results on the SPIE-AAPM-NCI BreastPathQ dataset.
\end{itemize}
\begin{figure}
	\centering
	\includegraphics[height=6.845cm,width=12cm]{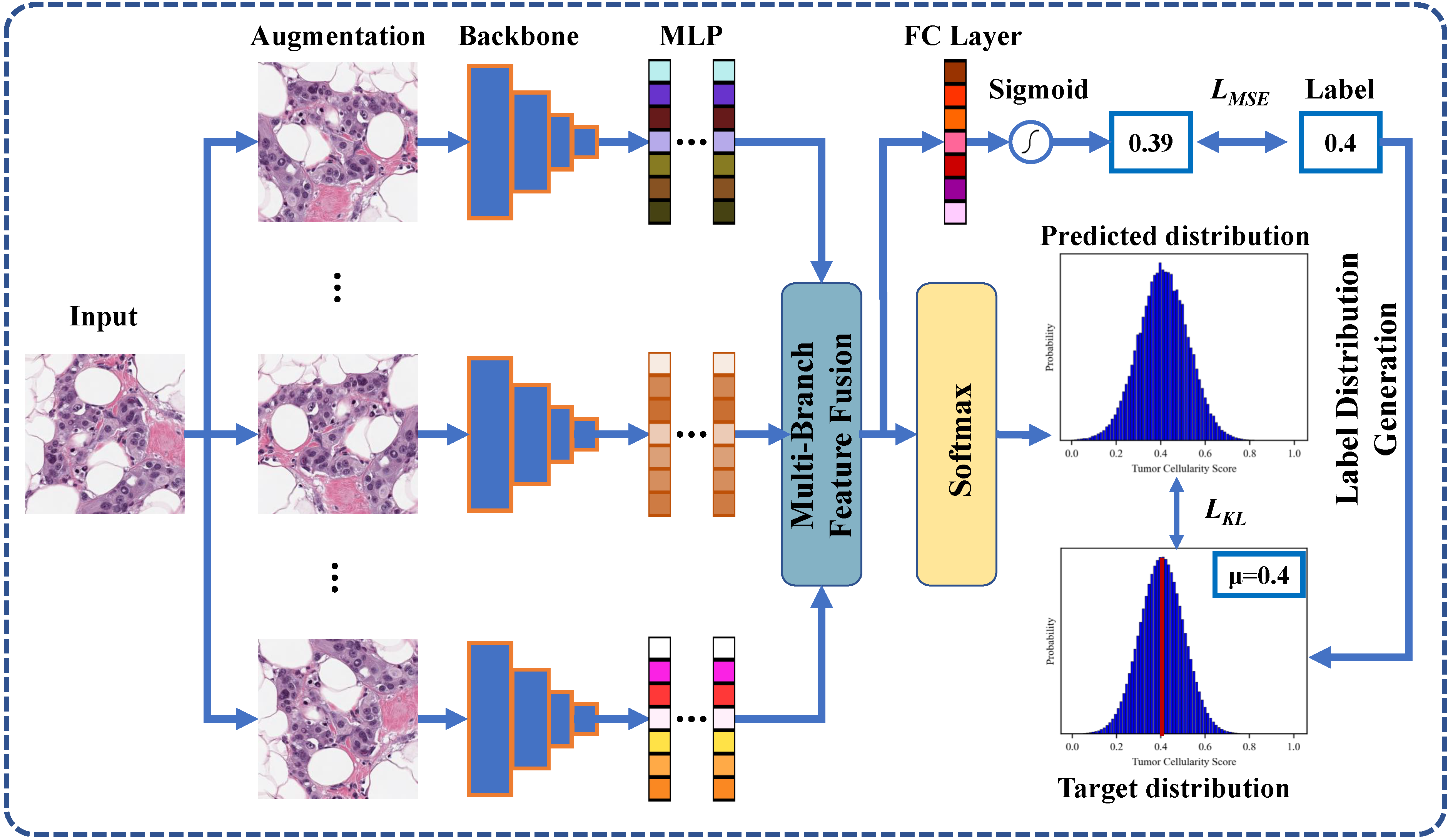}
	\caption[]{The overview of the proposed ULTRA. The input image patches are first augmented and fed into the backbone network to extract features. Furthermore, the feature maps are processed by MLPs in different branches and fused with multi-branch feature fusion. Finally, the network is optimized by jointly performing TC distribution learning and score regression.} 
	\label{fig:ultra}
\end{figure} 
\section{Methods}
The ULTRA aims to effectively exploit the uncertainty of the labels in breast tumor cellularity tasks by modeling the label ambiguity with a label distribution rather than a single value.
As is illustrated in Fig.\ref{fig:ultra}, it mainly consists of three parts: label distribution generation, multi-branch feature fusion, and label distribution learning. The details of each module are as follows:
\subsection{Label Distribution Generation}\label{subsec:ldg}
Following \cite{gao2017deep,tang2020uncertainty}, we first converted the single-value TC labels to discrete label distributions,  which effectively models the ambiguity among all possible TC labels. The TC distribution comprises a group of probability values which represent the description degree of each TC value for the input image. Specifically, to generate label distributions based on the TC scores, we first quantize the TC score ranges $\left[0,1\right]$ into an ordered label set $t\in \{t_{1}, t_{2},t_{i},...t_{n}\}$, where $n$ is the number of discretized labels, $t_{i}\in\left[0,1\right]$ are possible TC scores. We set $n=100$ according to the precision of TC labels. Given a TC value $s$ of a specific image, we then constructed a Gaussian distribution with label set $t$:
\begin{equation}
	G\left(t_{i} \mid s, \sigma\right)=\frac{1}{\sqrt{2 \pi} \sigma} \exp \left(-\frac{(t_{i}-s)^{2}}{2 \sigma^{2}}\right)
\end{equation}
where $s$ is the TC label and also the mean value of the Gaussian distribution.
$\sigma$ is the hyper-parameter that defines the sharpness of the Gaussian distribution and can be also taken as the uncertainty of TC score. We set $\sigma=0.04$ in our implementation since it achieved better results than other settings. Considering that a label distribution ${y_{i}}, i\in \{1,2,...n\}$ should satisfy two constrains, i.e., $y_{i} \in[0,1]$, and $\sum_{i} y_{i}=1$, we finally generate the TC label distribution by normalizing the Gaussian distribution:
\begin{equation}
	y_{i}=\frac{G\left(t_{i} \mid s, \sigma\right)}{\sum_{k} G\left(t_{k} \mid s, \sigma\right)}
\end{equation}

\subsection{Multi-Branch Feature Fusion}\label{subsec:MBFF}
To further explore the TC label uncertainties caused by multiple raters, we mimicked the multi-rater fusion process with a Multi-Branch Feature Fusion module (MBFF). The proposed MBFF significantly improved the TC estimation performance by effectively modeling the estimation process of multiple raters and leveraging the ambiguity among them. Specifically, given an input image $I$, we designed multiple branches to process it, and each branch represents the annotation process of a single rater. In a specific branch, we first applied augmentation methods to increase the variabilities of the input and further improve the generalization performance of the network; $\hat{I}_{k}$ denotes the processed image with augmentations at branch $k$. Then, the augmented image was further processed by a backbone network and an MLP in each branch, which modeled the decision process for a single rater. Finally, each branch generated various rich hierarchical features, which encoded valuable information for the annotation of the input image:
\begin{equation}
	f^{k} =  \mathcal{F}_{MLP}\left(\mathcal{F}_{Backbone}\left(\hat{I}_{k}\right)\right)
\end{equation}
where $\mathcal{F}_{MLP}$ and $\mathcal{F}_{Backbone}$ represent the MLP and the backbone network. $f^{k}$ are the feature maps at branch $k$. We used ResNet34 as the backbone network and three fully connected layers as the MLP in our implementation.

%Noted that all the backbone network and the MLP share parameters with each other to avoid too many parameters. Besides, 
Furthermore, to better exploit the uncertainties among different raters, we introduced MBFF which calculated the weighted average among features from different branches. Finally, we achieved the enhanced feature maps as follows:
\begin{equation}
	f_{enhanced} =  \frac{1}{N}\sum_{k=1}^{N}W_{k}f^{k}
\end{equation}
where $W_{k}$ is the relative importance of different branch, in our implementation, we set $W_{k}=1$ in all branches, empirically.

\subsection{Label Distribution Learning}\label{subsec:LDL}
The proposed ULTRA transfers the traditional TC regression problem to a label distribution learning problem (LDL), which better supervises the model to leverage the ambiguity of TC labels. 
%Most of the current methods take TC estimation as a simple regression problem, which ignores the ambiguity of TC labels, to fully exploit the label uncertainty of TC estimation, inspired by \cite{gao2017deep}, we proposed to perform label distribution learning for TC estimation. 
Based on the hierarchical features generated from the MBFF, the predicted TC distribution can be achieved by applying a softmax activation function:
\begin{equation}
	p_{i}=\frac{\exp \left(f^{i}_{enhanced}\right)}{\sum_{i} \exp \left(f^{i}_{enhanced}\right)}, \quad i=0,1,2, \cdots, n
\end{equation}
where $f^{i}_{enhanced}$ denotes the $i$th channel of enhanced feature from MBFF. 
Finally, we minimize the Kullback-Leibler (KL) divergence between the estimated
TC distribution $p_{i}$ and the normalized Gaussian distribution $y_{i}$:
\begin{equation}
	\mathcal{L}_{KL}=K L\left\{\boldsymbol{p}_{i} \| \mathbf{y_{i}}\right\}=\sum_{i} p_{i} \log \frac{p_{i}}{y_{i}}
\end{equation}

Moreover, existing label distribution learning suffers from severe object mismatch problem \cite{wang2021label, wang2021re}. Therefore, different from existing methods in \cite{gao2017deep,tang2020uncertainty}, which simply supervise the model with the KL divergence loss, we added an extra regression branch with mean square error loss function to mitigate the object mismatch problem, and finally jointly performed TC distribution learning and TC score regression in a multi-task learning manner:
\begin{equation}
	\mathcal{L}= \mathcal{L}_{KL}+ \alpha\mathcal{L}_{MSE}
\end{equation}
where $\alpha$ represents the relative weight to balance the importance of label distribution learning and TC score regression. Specifically, we empirically set $\alpha=1$ in our experiments. 
$\mathcal{L}_{MSE}$ denotes the mean square error between predicted TC score and labels.
%$\mathcal{L}_{MSE}=\frac{1}{M}\sum_{j=1}^{M}\left\|s_{j}-\hat{s}_{j}\right\|^{2}_{2}$ is the mean square error, where $\hat{s}_{j}$ is the regression score after sigmoid function for subject $j$, $s_{j}$ is the corresponding TC label, $M$ is the total number of subjects in a batch.

\section {Experiments and Results}
\subsection{Materials and Evaluation Metrics}
To demonstrate the effectiveness of the proposed method, we utilized the public SPIE-AAPM-NCI BreastPathQ \footnote{\url{https://breastpathq.grand-challenge.org/}} dataset\cite{petrick2021spie} in our study. 
%The dataset was collected from Sunnybrook Health Sciences Centre, Toronto, Canada, and it consists of 69 WSI scans of hematoxylin and eosin (H\&E)-stained slides of resection specimens from post-NAT patients with invasive residual breast cancer. The slides were scanned at 20$\times$ magnification (0.5 $\mu$m/pixel) using an Aperio AT Turbo 1757 scanner. 
The dataset provides 2394 and 185 patches with TC labels ranging from $0\%$ to $100\%$ in the training and validation set, and 1119 patches for testing whose TC labels are unavailable.

Following \cite{rakhlin2019breast}, we adopted intra-class correlation (ICC) \cite{shrout1979intraclass}, Cohen's kappa (Kappa)\cite{mchugh2012interrater} and mean square error (MSE) as the evaluation metrics.
%Noted that the Cohen's kappa is evaluated
%by binning the TC values into four classes of $0–25\%, 26–50\%,
%51–75\%$, and $76–100\%$.
\subsection{Implementation Details}
Our model was trained on the NVIDIA RTX 2080Ti GPU for 150 epochs. In the training phase, we first normalized the input data by subtracting the mean and dividing it by the standard deviation. In addition, we randomly performed different augmentations in MBFF including horizontal, vertical flips, elastic transforms, and etc. Besides, we utilized the Adam optimizer and set the initial learning rate to 1e-4, and the learning rate decayed by multiplying 0.1 every 100 epochs. We set the batch size to 8 in all our experiments empirically. We performed a two-stage training scheme: first stage for backbone training and the second stage for the whole framework training. In the testing phase, the input patches are also randomly augmented to perform multi-branch feature fusion, and then the network predicts the TC distributions for input patches. After obtaining the predicted TC distribution by performing the softmax function, the TC assessment is obtained by selecting the TC value with the largest probability among all possible TC scores. Finally, the predicted TC score is achieved by averaging the predictions from regression and label distribution branches.
\subsection{Experimental Results and Discussion}
%To demonstrate the effectiveness of the proposed ULTRA, we first performed several ablation studies to analyze the effectiveness of ULTRA. The experimental results are illustrated in Table \ref{tab:ablation_backbone}, \ref{tab:ablation_loss}. Then we further compared the ULTRA with state-of-the-art TC estimation methods, and the experimental results are illustrated in Table \ref{tab:SOTA_breastQ} and Fig. \ref{fig:wsi}. The detailed discussions are as follows: 
\textbf{Effectiveness of the MBFF:}\label{subsec:mb}
Table \ref{tab:ablation_backbone} shows the ablation study on the MBFF module. We tested three ULTRA variations with different numbers of branches, $N=\{1,2,3\}$. Experimental results demonstrate that the ULTRA with MBFF module outperforms its variants without fusion (i.e., N=1). This further proves the effectiveness of the MBFF. 
%\begin{minipage}{\textwidth}
%	\begin{minipage}[t]{0.48\textwidth}
%		\makeatletter\def\@captype{table}
%		\caption{Ablations of the Multi-branch Feature Fusion module, $N$ denotes the number of branches.}
%		\setlength{\tabcolsep}{0.2mm}
%		\begin{tabular}{cccc}
%			\toprule
%			Model     &ICC	&Kappa	&MSE\\
%			\midrule
%			ULTRA (N=1)  &0.919	&0.688	&0.013\\
%			ULTRA (N=2)      &0.921	&0.693	&0.014\\
%			ULTRA (N=3)  &\textbf{0.941}	&\textbf{0.703}	&\textbf{0.011}\\
%			\bottomrule
%		\end{tabular}	
%		\label{tab:ablation_backbone}
%	\end{minipage}
%	%\hfill
%	\begin{minipage}[t]{0.48\textwidth}
%		\makeatletter\def\@captype{table}
%		\caption{Ablations of the Label distribution Learning and Regression Branch for the ULTRA}
%		\setlength{\tabcolsep}{0.2mm}
%		\begin{tabular}{cccc}
%			\toprule
%			Model     &ICC	&Kappa	&MSE \\
%			\midrule
%			ULTRA w/o KL  &0.918	&0.600	&0.012 \\
%			ULTRA w/o MSE     &0.926	&0.650	&0.014 \\
%			ULTRA    &\textbf{0.941}	&\textbf{0.703}	&\textbf{0.011}\\
%			\bottomrule
%		\end{tabular}
%		\label{tab:ablation_loss}
%	\end{minipage}
%\end{minipage}
\begin{table}
	\parbox{.45\linewidth}{
		\centering
		\setlength{\tabcolsep}{1mm}
		\caption{Ablations of the MBFF, $N$: number of branches. $\uparrow$: The larger, the better; $\downarrow$: The smaller, the better.}
		\begin{tabular}{cccc}
			\toprule
			Model     &ICC$\uparrow$	&Kappa$\uparrow$	&MSE$\downarrow$\\
			\midrule
			ULTRA (N=1)  &0.919	&0.688	&0.013\\
			ULTRA (N=2)      &0.921	&0.693	&0.014\\
			ULTRA (N=3)  &\textbf{0.941}	&\textbf{0.703}	&\textbf{0.011}\\
			\bottomrule
		\end{tabular}	
		\label{tab:ablation_backbone}
	}
	\hfill
	\parbox{.45\linewidth}{
		\centering
		\setlength{\tabcolsep}{0.5mm}
		\caption{Ablations of the LDL and regression for the ULTRA. $\uparrow$: The larger, the better; $\downarrow$: The smaller, the better.}
		\begin{tabular}{cccc}
			\toprule
			Model    &ICC$\uparrow$	&Kappa$\uparrow$	&MSE$\downarrow$ \\
			\midrule
			ULTRA w/o KL  &0.918	&0.600	&0.012 \\
			ULTRA w/o MSE     &0.926	&0.650	&0.014 \\
			ULTRA    &\textbf{0.941}	&\textbf{0.703}	&\textbf{0.011}\\
			\bottomrule
		\end{tabular}
		\label{tab:ablation_loss}
	}
\end{table}

\textbf{Effectiveness of The LDL:}
%The main difference between the proposed ULTRA and other regression-based methods on TC estimation is the LDL. Therefore, we performed several experiments which kept either the regression or the label distribution learning branch. 
Table \ref{tab:ablation_loss} illustrates the experimental results in different settings. It demonstrates that the ULTRA with only LDL branch outperforms that of regression branch in ICC and Kappa, while slightly inferior in MSE. Moreover, the above two variants are inferior to the normal ULTRA in all metrics. This proves that the LDL effectively leveraged the label uncertainty and improved the TC estimation results. More importantly, the proposed ULTRA further enhanced the performance by combining the LDL and TC regression. 
\begin{figure}[htbp]
	\centering
	\subfigure[Results on different Standard Deviation of Gaussian distribution]{
		\includegraphics[width=5cm]{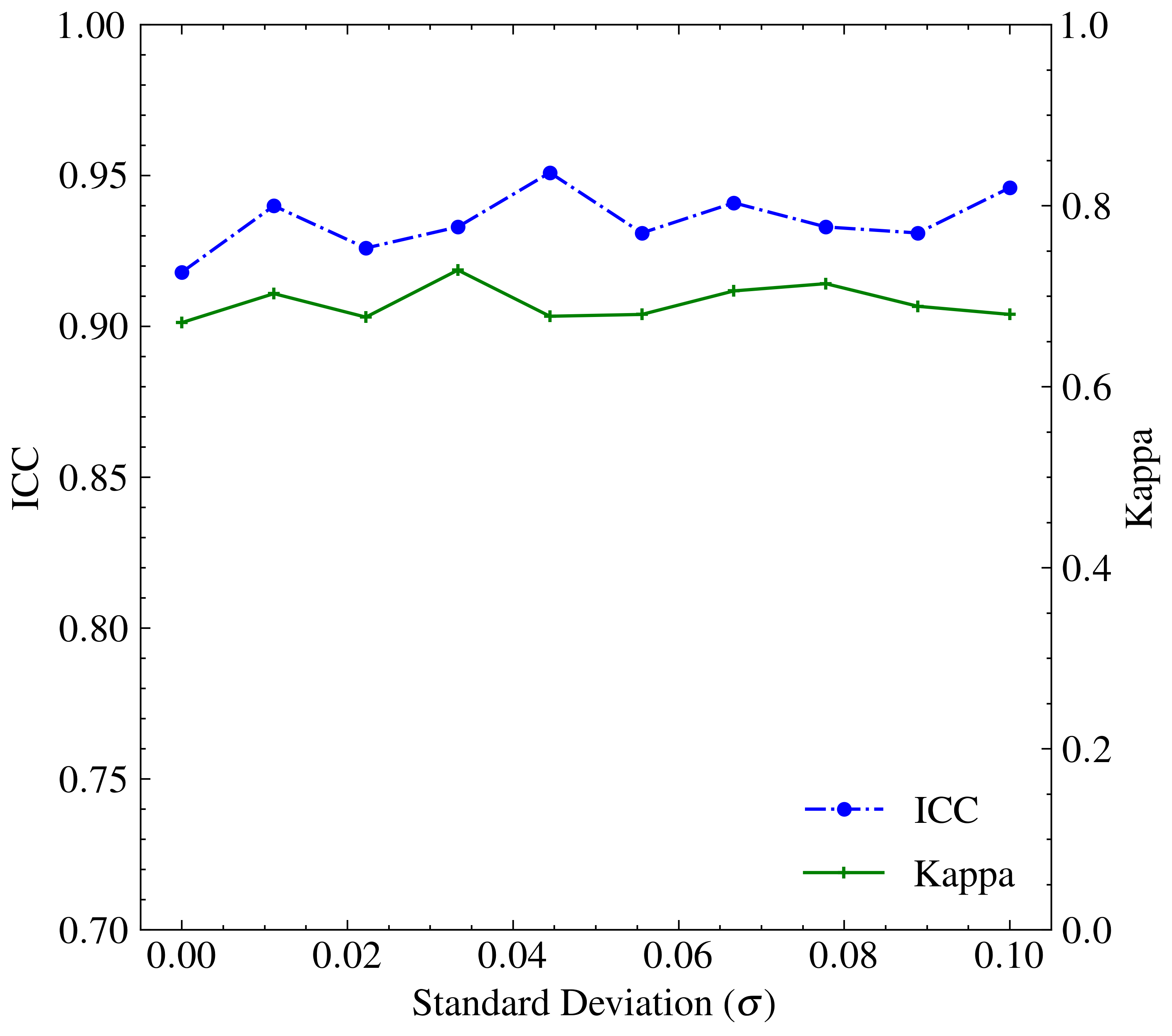}
		\label{subfig:sd}
	}
	\subfigure[Results on different number of branches]{
		\includegraphics[width=5cm]{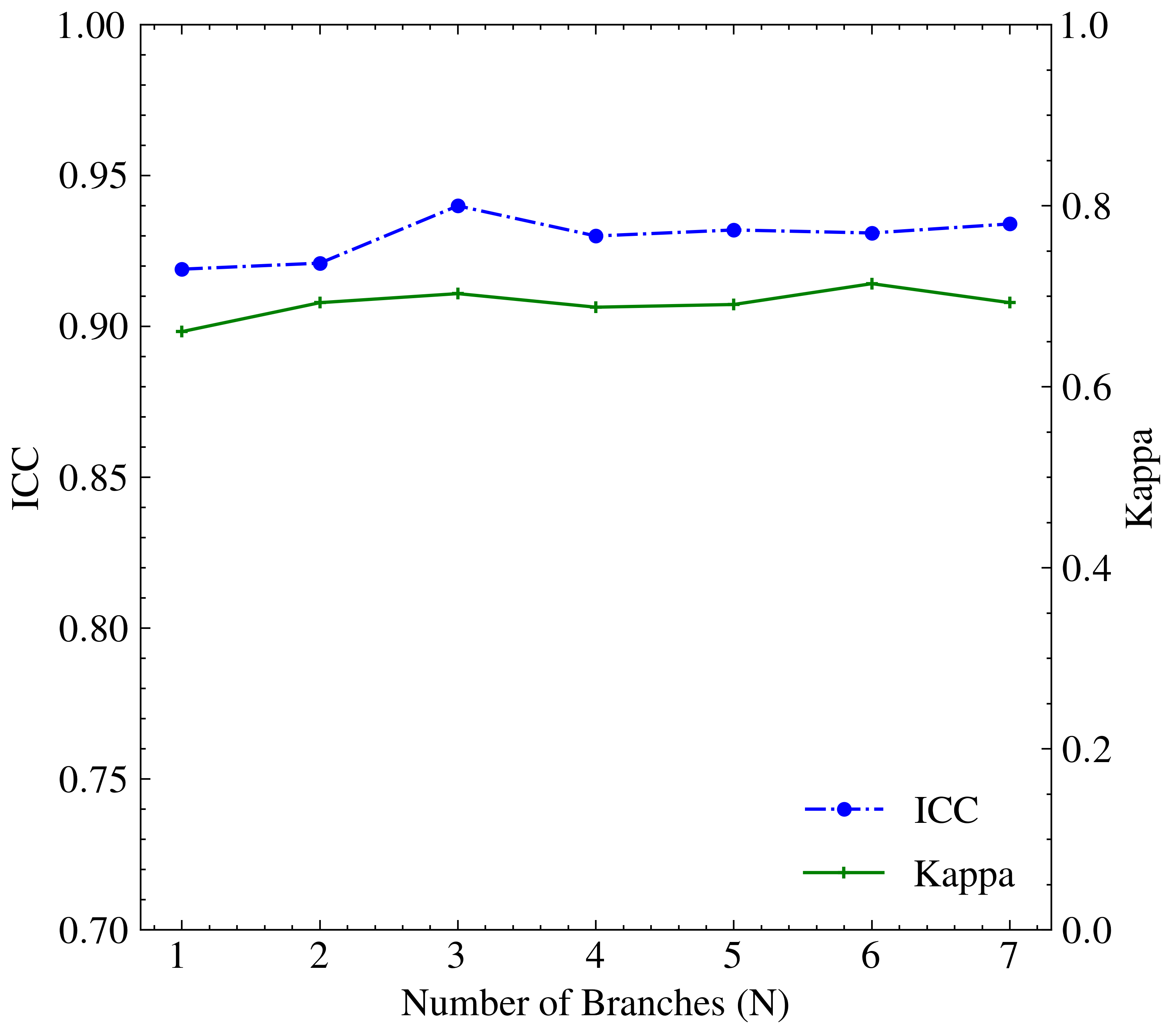}
		\label{subfig:num_branches}
	}
	%	\quad    %用 \quad 来换行
	%	\subfigure[SubCaption_3]{
	%		\includegraphics[width=2.5in]{PrintScreen_0003}
	%	}
	%	\subfigure[SubCaption_4]{
	%		\includegraphics[width=2.5in]{PrintScreen_0004}
	%	}
	\caption{The experimental results on different hyper-parameter settings}
	\label{fig:ablation}
\end{figure}
%\begin{figure}
%	\centering
%	\includegraphics[height=5cm,width=10.8cm]{images/SD_N.png}
%	\caption[]{The experimental results on different hyper-parameter settings} 
%	\label{fig:ablation}
%\end{figure} 
\begin{table}
	\centering	
	\caption{Quantitative results (Mean with $95\%$ confidence intervals) of state-of-the-art methods on BreastPathQ validation set. Bold text denotes the best result for that column. "-" means the authors didn't report that metric.}	
	\label{tab:SOTA_breastQ}
	\setlength{\tabcolsep}{1.5mm} % 此处将表格按照页面宽度调整		
	{\begin{tabular}[]{|l|c|c|c|}			
			%\toprule
			\hline	
			\multirow{2}{*}{Methods} &\multicolumn{3}{|c|}{Metrics}\\
			\cline{2-4}
			&ICC$\uparrow$  &Kappa$\uparrow$  &MSE$\downarrow$ \\	
			\hline	
			Baseline &0.901[0.870,0.930]	&0.688[0.602,0.774]	&0.015[0.011,0.019] \\
			Peikari et al.\cite{peikari2017automatic} &0.750[0.710,0.790] 	&0.380-0.420 &-\\
			Akbar et al.\cite{akbar2019automated} &0.830[0.790,0.860] &-	&-	\\
			Rakhlin et al.\cite{rakhlin2019breast}&0.883[0.858,0,905] 	&0.689[0.642,0,734] &\textbf{0.010[0.009,0.012]}\\
			Ours(ULTRA) &\textbf{0.941[0.920,0.950]}	&\textbf{0.703[0.620,0.787]}	&0.011[0.007,0.014]\\
			%\bottomrule	
			\hline		
	\end{tabular}}			
\end{table}

\textbf{Different Hyper-parameter Settings:}
We performed experiments to investigate the impact of two significant hyper-parameter settings: (1) Different $\sigma$ for Gaussian distribution. Following \cite{gao2017deep}, we uniformly sampled 10 values for $\sigma$ in range $\left[0,0.1\right]$. Fig.\ref{subfig:sd} illustrates the results for different $\sigma$. It shows that $\sigma$ should be set neither too large nor too small. This is reasonable since $\sigma$ controls the uncertainty of the TC label, large $\sigma$ would introduce extra label noise, while small $\sigma$ would not be enough to represent the ambiguity. (2) Different number of branches. Except for $N=\{1,2,3\}$, we tested more settings on the number of branches. Fig.\ref{subfig:num_branches} shows the results for different branches. The experimental results first improve and then decrease, finally tend to stable with the improvement of branches. $N=3$ achieves the best results.
 
\begin{figure}
	\centering
	\includegraphics[height=6.43cm,width=11cm]{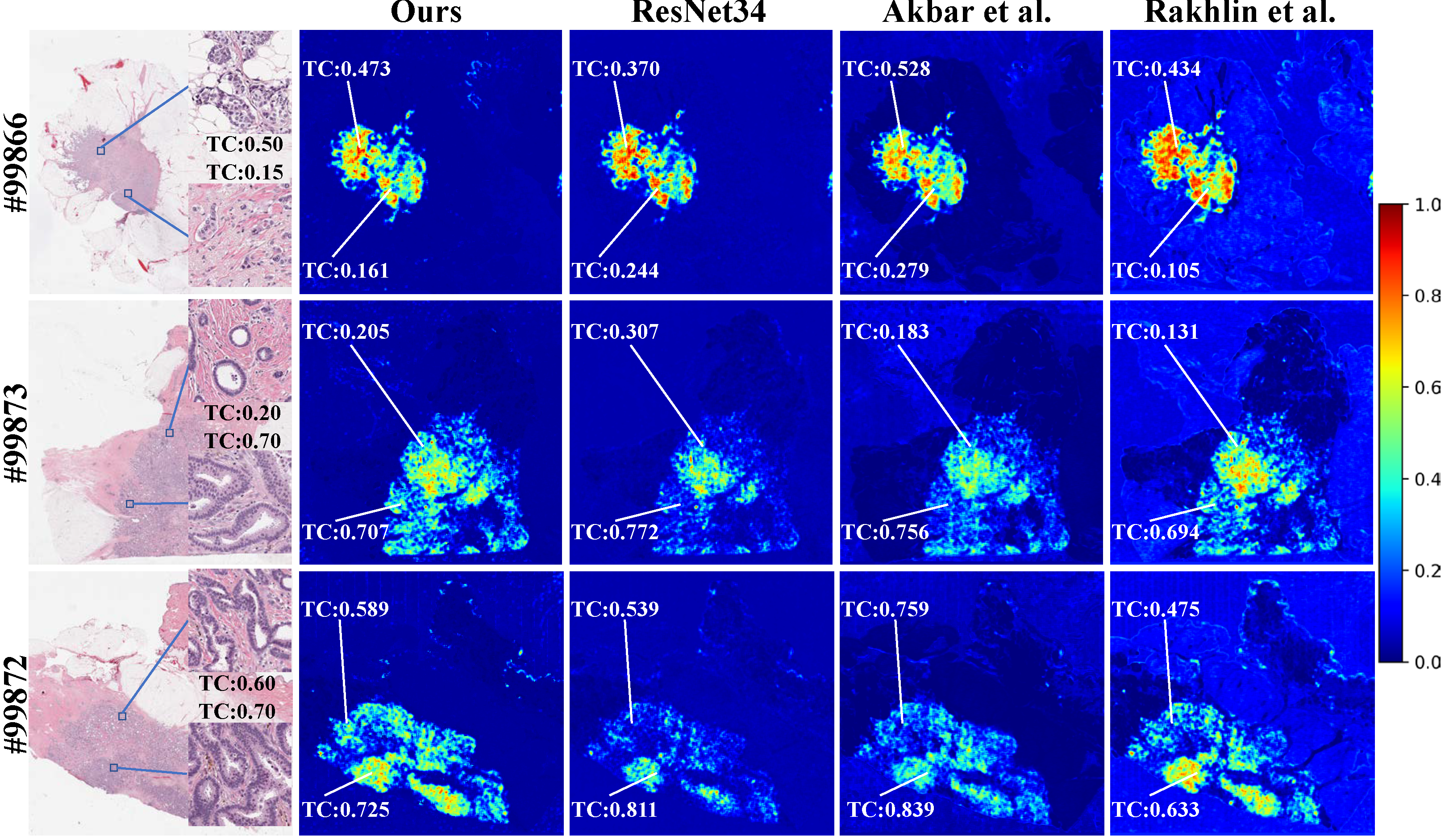}
	\caption[]{TC scores generated on WSIs of the BreastPathQ validation set.The blue color denotes healthy tissue (TC=$0\%$) and red denotes malignant (TC = $100\%$).} 
	\label{fig:wsi}
	%	 From left to right, it denotes the original WSIs, the TC scores produced by the proposed ULTRA, ResNet34, Akbar et al. and Rakhlin et al., respectively. 
\end{figure}

\textbf{Comparing with State-of-the-art Methods:}
We compared the experimental results with some SOTA methods including Peikari et al.\cite{peikari2017automatic}, Akbar et al.\cite{akbar2019automated}, and Rakhlin et al.\cite{rakhlin2019breast}. We also compared the results of a baseline network that performed direct regression with ResNet34. The proposed ULTRA takes the ResNet34 as the backbone network to ensure a fair comparison. Table \ref{tab:SOTA_breastQ} shows the experimental results on the BreastPathQ validation set. It demonstrates that the proposed ULTRA outperforms all other state-of-the-art methods on both ICC and Kappa metrics for a large margin and achieved comparable results on the MSE metric. The superior TC estimation results further prove the effectiveness of the proposed method. In addition, we performed t-test between the ULTRA and other SOTA methods, the experimental results prove that
the superiority of the ULTRA is statistically significant (p-value<0.001,p-value<0.005, p-value<0.01 for ICC, Kappa and MSE, respectively).
Moreover, we also compared the corresponding TC scores generated by each method on WSIs of the BreastPathQ validation set, which is illustrated in Fig.\ref{fig:wsi}. It demonstrates that the TC scores generated from the proposed methods are closer to the ground-truth labels than other techniques, proving that the proposed method can effectively assess breast tumor cellularity. 
\section {Conclusion}
In this paper, we proposed ULTRA to address the problem of ignoring label ambiguity for the tumor cellularity assessment task. The proposed ULTRA transformed TC regression to a label distribution learning problem, which significantly exploited the label ambiguity of the TC scores. Moreover, by mimicking the multi-rater fusion process in clinical routines, the framework further leveraged the label uncertainty and improved the TC estimation performance. Experimental results prove that the proposed ULTRA achieved superior performance compared to many state-of-the-art methods. 
\section*{Acknowledgments}
This work was supported by the National Natural Science Foundation of China under Grant Grant 62001144 and 62001141, and by Science and Technology Innovation Committee of Shenzhen Municipality under Grant  JCYJ20210324131800002 and RCBS20210609103820029.
\bibliographystyle{splncs}
\bibliography{reference}

\end{document}